\documentclass[10pt,conference]{IEEEtran}

\usepackage[backend=biber,style=ieee]{biblatex} 
\usepackage{amsmath,amssymb,amsfonts}
\usepackage{algorithmic}
\usepackage{graphicx}
\usepackage{textcomp}
\usepackage{xcolor}
\usepackage{balance}
\usepackage{listings}
\usepackage{xspace}

\usepackage[hidelinks,bookmarksdepth=-2]{hyperref}

\input{macros}

\begin{document}

\title{Escaping the Time Pit: Pitfalls and Guidelines for Using Time-Based Git Data}

\author{\IEEEauthorblockN{Samuel W. Flint}
\IEEEauthorblockA{\textit{University of Nebraska--Lincoln}\\
Lincoln, NE, USA\\
swflint@huskers.unl.edu}
\and
\IEEEauthorblockN{Jigyasa Chauhan}
\IEEEauthorblockA{\textit{University of Nebraska--Lincoln}\\
Lincoln, NE, USA\\
jchauhan2@huskers.unl.edu}
\and
\IEEEauthorblockN{Robert Dyer}
\IEEEauthorblockA{\textit{University of Nebraska--Lincoln}\\
Lincoln, NE, USA\\
rdyer@unl.edu}
}

\maketitle

\begin{abstract}
Many software engineering research papers rely on time-based data (e.g., commit timestamps, issue report creation/update/close dates, release dates).
Like most real-world data however, time-based data is often dirty.  To date, there are no studies that quantify how frequently such data is used by the software engineering research community, or investigate sources of and quantify how often such data is dirty.
Depending on the research task and method used, including such dirty data could affect the research results.
This paper presents the first survey of papers that utilize time-based data, published in the Mining Software Repositories (MSR) conference series.  Out of the 690 technical track and data papers published in MSR 2004--2020, we saw at least 35\% of papers utilized time-based data.  We then used the Boa and Software Heritage infrastructures to help identify and quantify several sources of dirty commit timestamp data.  Finally we provide guidelines/best practices for researchers utilizing time-based data from Git repositories.

\end{abstract}


\section{Introduction}
\label{sec:intro}

The Mining Software Repositories (MSR) conference has been around as a workshop, working conference, and finally a full conference since 2004.  During those 17 years there have been almost 600 research and over 100 data papers published.  The majority of the research in MSR relies on analyzing existing data, including data from version control systems (CVS, Subversion, Git), issue/bug reports, discussion forums (emails, Stack Overflow), pull requests (PRs), continuous build/test systems, etc.  Often these data sources include time components indicating when events occurred, such as the timestamp of a code commit or when a pull request was opened or closed.

Depending on the source of the data, there may be errors or inconsistencies in the time components.  For example, the Git version control system (VCS) allows users to specify both the authored and committed dates when creating new commits. It also allows editing the existing commit graph (rebasing) which allows for modification of the timestamps of older commits.  There are also more general issues with time data, for example dealing with inconsistent time zones and clock skews.

To date, no survey has been performed to investigate how MSR researchers utilize time-based data in their research.  This work thus surveys 690 MSR technical research and data showcase papers from 2004--2020 to determine how many rely on time-based data and what techniques are utilized to control for potential errors in that data.  We utilize keyword searches of the papers and then manual inspection to determine that at least 169 technical research papers and 70 data showcase papers rely on or provide time-based data.  This accounts for at least 35\% of the papers in MSR's history.

Based on the survey results indicating that VCS is the most used data kind incorporating time-based data and that GitHub is the most used data source, we investigate potential problems with time-based Git data.  We utilize the Boa~\cite{boa,boa-website} and Software Heritage~\cite{software-heritage,software-heritage-archive} infrastructures and attempt to quantify how frequently some kinds of errors occur.  We also attempt to infer the potential source(s) of these errors.  Based on this we can see a couple of potential pitfalls when utilizing time-based data, and how frequently one might encounter them.

The results show that over 4k commits have timestamps that are suspiciously too old (even before the initial release of CVS, 19 November 1990).  Many of those bad timestamps were the result of tools such as \texttt{git-svn}.  We also discovered over 18k commits from over 4k projects where one (or more) of the commit's parent commits had a timestamp that was newer than the commit itself --- something that does not make sense.  Again, many of these were the result of automated tools or a small set of users.  A replication package containing all of the data and scripts used in our analysis is also publicly available~\cite{replication}.

Finally we propose some guidelines for researchers utilizing time-based Git data to help escape these pitfalls.  These include filtering out older projects (based on our analysis, we would recommend anything from 2013 or older), filtering out certain projects or users that seem to have a lot of bad commit timestamps, or preferably running specific analyses to automatically verify and reject commits with suspicious timestamps.  
We hope future MSR researchers follow these guidelines.

In the next section we discuss prior surveys of MSR research.  In \secref{sec:survey} we detail our survey on the use of time-based data in MSR research.  Then in \secref{sec:problems} we attempt to identify and quantify some examples of problems with time-based data in Git.  We discuss implications of the survey and present best practice guidelines in \secref{sec:discussion} and finally conclude in \secref{sec:conclusion}.

\begin{table*}
\caption{Published and selected MSR papers, by year. Due to space, we omit the full list of selected papers and refer to the Excel spreadsheet in our replication package~\cite{replication}.}
\label{tab:years}
\newcommand{\oldtabcolsep}{\tabcolsep}
\renewcommand{\tabcolsep}{4.6pt}
\begin{tabular}{r|c|c|c|c|c|c|c|c|c||c|c|c|c|c|c|c|c||c}
\rowcolor{gray!20}
\textbf{year} & \textbf{2004} & \textbf{2005} & \textbf{2006} & \textbf{2007} & \textbf{2008} & \textbf{2009} & \textbf{2010} & \textbf{2011} & \textbf{2012} & \textbf{2013} & \textbf{2014} & \textbf{2015} & \textbf{2016} & \textbf{2017} & \textbf{2018} & \textbf{2019} & \textbf{2020} & \textbf{total} \\
\hline
\hline
\textbf{technical} & 26 & 22 & 28$^*$ & 27 & 31 & 23 & 22 & 27 & 31 & 37 & 39 & 42 & 42$^+$ & 43 & 48 & 47 & 45 & \textbf{580} \\
\textbf{selected}  &  5 &  2 &      6 &  8 &  7 &  9 &  6 &  9 &  9 & 10 & 13 & 12 &     11 & 13 & 13 & 16 & 20 & \textbf{169} \\
\hline
\textbf{data}      & -- & -- &     -- & -- & -- & -- & -- & -- & -- & 14 & 15 & 16 &      7 &  7 & 15 & 17 & 19 & \textbf{110} \\
\textbf{selected}  & -- & -- &     -- & -- & -- & -- & -- & -- & -- &  9 & 12 & 13 &      4 &  6 & 10 &  2 & 14 &  \textbf{70} \\
\hline
& 19\% & 9\% & 21\% & 30\% & 23\% & 39\% & 27\% & 33\% & 29\% & 37\% & 46\% & 43\% & 31\% & 38\% & 37\% & 28\% & 53\% & \textbf{35\%} \\
\end{tabular}
\renewcommand{\tabcolsep}{\oldtabcolsep}
\centering
{\scriptsize
($^*$2006 had 2 papers listed in the program that do not appear in the proceedings, and we excluded them)\\
($^+$2016 had 1 paper not listed in the program that appears in the proceedings, and we included it)}

\end{table*}

\section{Previous Studies}
\label{sec:related}






In this section we discuss prior works that either performed surveys of MSR research or propose guidelines for future MSR researchers to follow.


\textcite{demeyer13} explored 10 years of MSR papers to determine what software projects were studied and the frequency of studies on the given projects, as well as the infrastructure behind mining.  In particular, they noted that the most common source of data were version control systems, including the then-building popularity of Git, and infrequency of use of VCS other than CVS, Subversion or Git.  They also noted that few of the studies at the time had considered industrial cases and instead most were over open source software.

\textcite{perils,perils2} addressed various characteristics of GitHub repositories. They note several possible problems with GitHub data, such as containing personal and inactive projects or that many pull requests are not marked as merged despite being so. They provide guidelines for software engineering researchers on how to use GitHub data in their research more effectively.

\textcite{cosentino16} reviewed the use of GitHub data in prior studies and structured data archives.  In particular, they looked at how GitHub data was used, how the data was collected, and what, if any, limitations were reported in the studies.  The operation of the GitHub API at the time, particularly in terms of request limits and inconsistent responses, was noted as a limitation.  Further, the lack of availability of fresh data was considered as a potential issue, due to reliance on common curated data sources.  Finally, they also described potential issues with sampling of datasets, suggesting that better sampling methods are needed.

\textcite{robles10} was concerned with the replication of MSR studies.  Replication required the availability of datasets and tools, as well as an adequate description of techniques used to filter and analyze those datasets.  The tools and descriptions that are preserved for replication may filter using time, yet this particular class of filtering criteria is only one of many which must be considered for replication.

Like \cite{robles10}, \textcite{ghezzi13} studied replication of MSR studies.  In particular, they described a web service to gather and analyse software repository data, which was then used to replicate MSR studies from 2004--2011.  They found that, of the studies in those years, 51\% could not be fully replicated.

\textcite{kotti19} looked at the use of pre-prepared datasets, particularly those described in MSR's Data Showcase.  As these datasets often include a time component (see \secref{subsec:paper-selection}), their definitions and methods are important.  Further, we note that the use of these datasets presents an opportunity for time-based errors to propagate into further studies and incorporating filtering/cleaning techniques into such datasets is extremely important.

\textcite{hemmati13} described a set of best-practices, a ``cookbook'', for mining software repositories researchers.  This included suggestions regarding social data, statistical and analytical techniques, and the sharing of tools.  They discuss the issue of VCS noise and the potential lack of granularity in VCS-recorded changes, however, they do not discuss the potential causes of discontinuities in time data, nor ways they may be handled.  

\textcite{gasser04}, early on in MSR's history, evaluated the needs of researchers in the field, and the data and artifacts to be studied.  They proposed a set of characteristics for studies to have, and discussed issues with data and how these issues may be addressed.  In particular, they discussed the frequent need to \emph{normalize} data as part of the analysis and data collection process.

\textcite{bird09} discussed mining one VCS in particular, Git, and the potential issues that may occur in mining repositories using it.  This work describes a number of issues, in particular, the existence of rebasing, which allows users to rewrite a repository's history, re-using commits in a different order than the commit timestamps may suggest.  We consider this as a potential cause of some of the time problems we discover in \secref{sec:problems}.

\textcite{paixao19:rebasing} study how rebasing is used in code reviews.  In particular, they study what the side effects of rebasing are during the code review process.  Because of the side effects observed, specifically the introduction of commits not germane to the review, they present a methodology to better handle rebasing in mining code reviews.  Although rebasing causes issues, their study was not focused directly on the time changing aspects of rebasing.

\section{Survey on the Use of Time-Based Data}
\label{sec:survey}

This paper investigates the following research questions:

\begin{enumerate}[label={\textbf{RQ\arabic*}}]
    \item \textbf{How many MSR papers rely on time-based data?} We look at all MSR published technical and Data Showcase papers and use keyword searches to identify time-based data being used.

    \item \textbf{What kinds of data include time?} We classify the papers according to the data kinds.

    \item \textbf{What filtering or cleaning techniques are used with time-based data?} We classify the papers according to their filtering or cleaning techniques.

    \item \textbf{Is bad time-based data common?} Based on the results of the prior research questions, we investigate Git data from GitHub to quantify how frequently bad time-based data occurs.
\end{enumerate}

We begin by by surveying published MSR proceedings.  We select papers to review, then from these, classify what kind of time-related data is used, how it is filtered or cleaned, and the source of the data used.

\subsection{Paper Selection}
\label{subsec:paper-selection}

For this study we focused only on papers published in MSR proceedings from 2004 to 2020.  All technical track papers (short and long) and Data Showcase papers were considered.  Data Showcase papers were included as they are potential data sources for other (future) research papers.  Mining Challenge papers were excluded, as all papers in this category for a given year typically use the same challenge dataset, which may skew results towards a particular kind of data in that year.  This gave us a corpus of 690 papers to inspect.

Papers from this corpus were filtered by one author, retaining for further study those that contained any of the following time-related keywords:  \texttt{time}, \texttt{date}, \texttt{epoch}, \texttt{record}, \texttt{month}, \texttt{year}, \texttt{hour}, \texttt{minute}, \texttt{second}, \texttt{period}, \texttt{week}, \texttt{chronolog}, \texttt{day}, \texttt{past}, and \texttt{interval}.

After the papers were enumerated, two authors determined what kinds of time-based data was used, the source(s) of the data, and any methods used to filter, clean, or normalize the time-based data.  During this process, if any two authors both found that a paper did not fit the study, it was removed (in particular, the use of ``runtime'' as a performance metric or ``epoch'' as a measure of training time was considered irrelevant to this study).

Additionally, if there was disagreement on the kinds of data, source of data, or filtering techniques both authors discussed until agreement was reached.  This affected a total of 5 rows for source of data, 5 rows for kinds of data, and 22 rows for filtering techniques.  This data is described in more detail in the following subsections.

A total of 43 papers (6.2\%) had a matching keyword but were removed.  Thus the keyword search yielded a precision of 85\%.  The results of this selection process are shown in \tabref{tab:years} and the spreadsheet including all identified papers and human judgements is available in our replication package~\cite{replication}.


\subsection{Paper Classification}
\label{subsec:classification}

We found that across the 239 papers selected, 44 different kinds of time-including data were used.  From these, all data kinds used by more than one paper are shown in \tabref{tab:data-kinds}.  In particular, we found that VCS data (diffs, commit lineage, commit logs, authors, \textit{etc.}) are the most commonly used.

\begin{table}
\caption{Common kinds of data used in MSR papers. Only data kinds used by more than one paper are listed here.}
\label{tab:data-kinds}
\centering
\begin{tabular}{rl}
\hline
\textbf{Number of Papers} & \textbf{Data Kind} \\
\hline
153 (64\%) & version control systems (VCS) \\
 73 (31\%) & issues \\
 40 (17\%) & forge metadata \\
 39 (16\%) & releases \\
 16 \hspace{0.5 em}(7\%) & pull requests (PRs) \\
 15 \hspace{0.5 em}(6\%) & mailing lists \\
 12 \hspace{0.5 em}(5\%) & continuous improvement (CI) builds \\
 10 \hspace{0.5 em}(4\%) & Stack Overflow \\
 10 \hspace{0.5 em}(4\%) & system logs \\
  4 \hspace{0.5 em}(2\%) & Common Vulnerabilities and Exposures (CVE) \\
  2 \hspace{0.5 em}(1\%) & time cards \\
  2 \hspace{0.5 em}(1\%) & chat logs \\
\hline
\end{tabular}

\end{table}

Further, we found roughly 180 different sources of data were used (proprietary or custom repositories were grouped as a single source, `anonymized', rather than listed separately).  Data sources used by more than one paper are shown in \tabref{tab:data-sources}.  Perhaps unsurprisingly, GitHub is the most common data source.  Eclipse, Apache\footnote{Apache projects are listed separately if a paper studied just that single project and not the full Apache repository, i.e., because NetBeans is mentioned separately from other Apache projects, it is listed alone. Similarly, httpd and OpenOffice are listed separately.}, and Mozilla repositories are also popular, as is the GHTorrent dataset.

\begin{table}
\caption{Common data sources used by MSR papers. Only data sources used by more than one paper are listed here.}
\label{tab:data-sources}
\centering
\begin{tabular}{rl}
\hline
\textbf{Number of Papers} & \textbf{Data Source} \\
  \hline
49 (21\%) & GitHub \\
19 \hspace{0.5 em}(8\%) & Eclipse \\
19 \hspace{0.5 em}(8\%) & Anonymized \\
17 \hspace{0.5 em}(7\%) & Apache organization repositories \\
14 \hspace{0.5 em}(6\%) & GHTorrent \\
13 \hspace{0.5 em}(5\%) & Mozilla \\
9 \hspace{0.5 em}(4\%) & Stack Overflow \\
8 \hspace{0.5 em}(3\%) & Firefox \\
7 \hspace{0.5 em}(3\%) & PostgreSQL \\
7 \hspace{0.5 em}(3\%) & Linux \\
5 \hspace{0.5 em}(2\%) & Travis \\
5 \hspace{0.5 em}(2\%) & Python \\
5 \hspace{0.5 em}(2\%) & Debian \\
4 \hspace{0.5 em}(2\%) & OpenStack \\
4 \hspace{0.5 em}(2\%) & Google Play \\
3 \hspace{0.5 em}(1\%) & Spring \\
3 \hspace{0.5 em}(1\%) & Source Forge \\
3 \hspace{0.5 em}(1\%) & OpenOffice \\
3 \hspace{0.5 em}(1\%) & NetBeans \\
3 \hspace{0.5 em}(1\%) & Maven Central Repository \\
3 \hspace{0.5 em}(1\%) & JBoss \\
3 \hspace{0.5 em}(1\%) & EUDC \\
3 \hspace{0.5 em}(1\%) & Chromium \\
3 \hspace{0.5 em}(1\%) & BIRT \\
3 \hspace{0.5 em}(1\%) & ArgoUML \\
2 \hspace{0.5 em}(1\%) & World of Code \\
2 \hspace{0.5 em}(1\%) & Software Heritage Archive \\
2 \hspace{0.5 em}(1\%) & QT \\
2 \hspace{0.5 em}(1\%) & OpenBSD \\
2 \hspace{0.5 em}(1\%) & MySQL \\
2 \hspace{0.5 em}(1\%) & Mycomp \\
2 \hspace{0.5 em}(1\%) & Moodle \\
2 \hspace{0.5 em}(1\%) & LKML \\
2 \hspace{0.5 em}(1\%) & KDE \\
2 \hspace{0.5 em}(1\%) & JBOSS \\
3 \hspace{0.5 em}(1\%) & Apache httpd \\
2 \hspace{0.5 em}(1\%) & Google Code \\
2 \hspace{0.5 em}(1\%) & GHtorrent \\
2 \hspace{0.5 em}(1\%) & GCC \\
2 \hspace{0.5 em}(1\%) & F-Droid \\
2 \hspace{0.5 em}(1\%) & Evolution \\
2 \hspace{0.5 em}(1\%) & Eureka \\
2 \hspace{0.5 em}(1\%) & DockerHub \\
2 \hspace{0.5 em}(1\%) & CVE \\
2 \hspace{0.5 em}(1\%) & Chrome \\
2 \hspace{0.5 em}(1\%) & BugZilla \\
2 \hspace{0.5 em}(1\%) & App Stores \\
\hline
\end{tabular}

\end{table}

Finally we investigated any filtering or cleaning techniques used by the selected papers. There were 54 different methods of cleaning or filtering the data (considering all custom conditions as one method for the purposes of counting).  Any used by more than one paper are shown in \tabref{tab:filtering-methods}.  We further analyze these in the next section.

\begin{table}
\caption{Common filtering/cleaning techniques used by MSR papers. Only techniques used by more than one paper are listed here.}
\label{tab:filtering-methods}
\centering
\begin{tabular}{rrlc}
\hline
\textbf{Times Used} & \textbf{Method Used} & \textbf{Type}\\
\hline
 149 (62\%) & none explicitly mentioned & -- \\
  24 (10\%) & time window & filtering \\
  15 \hspace{0.5 em}(6\%) & date cutoff & filtering \\
   7 \hspace{0.5 em}(3\%) & custom condition & filtering \\
   5 \hspace{0.5 em}(2\%) & changeset coalescence & filtering \\
   4 \hspace{0.5 em}(2\%) & CVSAnalY & filtering \\
   3 \hspace{0.5 em}(1\%) & date format correction & cleaning \\
\hline
\end{tabular}

\end{table}

\subsection{Identified Filtering and Cleaning Techniques}
\label{sec:filtering}

Among the various time-based filtering and cleaning techniques found in MSR papers, we found six used by more than one paper.  The majority of these are filtering techniques of some form, with a single cleaning technique described.  It is important to note, however, the majority of papers utilizing time-based data (149, 62\%) do not explicitly describe any filtering or cleaning methods used.  We discuss each of the six techniques in more detail.

\subsubsection{time window}

A number of studies select data from a source that was added between two dates or other markers in time (\textit{e.g.,} releases).  This was by far the most common explicitly described method, being found in 24 studies.  Some of these studies provided full dates~\cite{durieux20,pimentel19}, others only partial dates out to year or month~\cite{hayashi19,ahasanuzzaman16}, or version numbers of releases~\cite{antoniol05}.

\subsubsection{date cutoff}

All studies retrieved data from before a specific date (the date of the study).  But whether the study date is used or some other date is used must be considered.  In particular, some papers describe what their cutoff date for data inclusion is, while others do not.  This method is used in particular by~\textcite{Wang_2020,Karampatsis_2020,Zhu_2019,Cito_2017}.

\subsubsection{custom condition}

A custom condition specifies some method for filtering a data source using time.  These were frequently employed to ensure that commits or issues were studied that matched some temporal condition relating the two, or to ensure that commits were in order, as well as for other purposes.

\textcite{liu2020} describe the use of a particular time-based condition to select commits to study.  They were interested in finding commits between the open and close of a particular issue (in other words, looking for fixing commits).  This condition is $issue_{create} < commit_{create} < issue_{close}$, and uses time components from both issues and commits.

\textcite{kikas16} use commit time and a forge's repository creation time to remove forks of original projects so that only the originals may be studied.  We note in particular that this method may inappropriately remove projects which have changed forges.

Finally, \textcite{steff12} construct a commit graph such that, for each commit node, it is only connected to nodes preceding it in time which also share files in common, that is, for two commits $(t_1, \mathcal{F}_1)$ and $(t_2, \mathcal{F}_2)$, $(t_1, \mathcal{F}_1) \to (t_2, \mathcal{F}_2)$ if and only if $t_1 < t_2$ and $\mathcal{F}_1 \cap \mathcal{F}_2 \neq \emptyset$.


\subsubsection{changeset coalescence}

Further techniques used include changeset coalescence or commit reconstruction.  This technique is useful in CVS or RCS repositories where changes are only made to individual files.  Most of these methods operate by collecting changes made in a small window (3 minute) by one user into a single changeset; they may also be aided by the use of ChangeLogs to collect such changes.  This technique was used by~\textcite{zimmerman04,walker06,kagdi06,dambros10}.

\subsubsection{CVSAnalY}

CVSAnalY\footnote{\url{https://github.com/MetricsGrimoire/CVSAnalY}} is a tool to extract information from VCS logs of various kinds.  It supports CVS, Subversion, and Git.  When operating on Subversion repositories, it skips over commits it considers invalid, with one condition being the lack of date.  Otherwise, it performs a sort of date format correction, storing all dates as Unix timestamps with associated time zones.

In particular, this tool sees use on Git repositories~\cite{gonzalez-barahona15,robles14,goeminne13}, as well as subversion repositories~\cite{sadowski11}, where the filtering may be most apparent.

\subsubsection{date format correction}

Due to the diversity of data sources and systems used, date and time data must often be normalized, that is, put into a standard format.  This may include time zone conversion or other actions, and presents a single, unified view of time for analysis and further filtering.  This cleaning technique is specifically used by~\textcite{claes20,xu18,baysal12}.





These are some of the most common time-based data filtering techniques used.  In the next section we investigate and attempt to quantify how frequently problems occur in time-based data.
\section{Identifying and Quantifying Time Problems}
\label{sec:problems}

In the previous section we showed that many MSR papers rely on time data in their methodologies.  In this section we show some potential problems with this kind of data and attempt to quantify how often these problems occur.

To aid in quantification, we utilize the Boa infrastructure~\cite{boa,boa-website}.  We rely on the latest general dataset (not datasets specific to a single language), which was ``2019 October/GitHub''.  This dataset contains 282,781 Git repositories from GitHub that each contain a minimum of one revision.\footnote{\url{http://boa.cs.iastate.edu/boa/?q=boa/job/public/90189}}  There are a total of 23,229,406 revisions in the dataset.\footnote{\url{http://boa.cs.iastate.edu/stats/index.php}}

\begin{figure}[ht]
\begin{Boa}
P:  output collection[string][string] of time;
P2: output collection[string][string] of time;
P3: output collection[string][string] of time;

cvs_release := T"Mon Nov 19 00:00:00 UTC 1990";
boa_dataset := T"Thu Oct 31 00:00:00 UTC 2019";
last: Revision;

visit(input, visitor {
	before r: Revision -> {
		if (r.commit_date < cvs_release)
			P[input.project_url][r.id] << r.commit_date;
		else if (r.commit_date > boa_dataset)
			P2[input.project_url][r.id] << r.commit_date;

		if (def(last)
		    && r.commit_date < last.commit_date
		    && !match("merge", lowercase(last.log))
		    && !match("merge", lowercase(r.log)))
			P3[input.project_url][r.id] << r.commit_date;
		last = r;
	}
});
\end{Boa}
  \caption{Boa query to find bad commit timestamps for this study.  This query is the combination of \url{http://boa.cs.iastate.edu/boa/?q=boa/job/public/90164}, \url{http://boa.cs.iastate.edu/boa/?q=boa/job/public/90169}, and \url{http://boa.cs.iastate.edu/boa/?q=boa/job/public/90973} for presentation purposes.}
  \label{fig:boaquery}
\end{figure}

\figref{fig:boaquery} shows the relevant Boa query used to collect data for this section.  Note that Boa stores commits for the main branch only, in a topologically sorted array based on the commit parent(s).  This means traversals on the commits (called \texttt{Revision}s in Boa) are performed in topological order.

The query looks for two possible kinds of bad time data.  First, it looks for suspicious commit timestamps that seem too old.  Then it also looks for commits who have a parent that is newer than themselves.  The query outputs the project URL, the commit ID, and the commit timestamp.

Since the notes for Boa indicate this data was actually from 2015\footnote{\url{http://boa.cs.iastate.edu/boa/?q=content/dataset-notes-october-2019}}, many projects no longer exist on GitHub.  We also utilized the Software Heritage archive~\cite{software-heritage,software-heritage-archive} to attempt to locate these repositories that have since been deleted.

In this section, we investigate these two sources of bad time data in more detail.

\subsection{Looking for Suspicious Commit Timestamps}
\label{subsec:old-commits}

First we investigated to see if there were suspicious commit timestamps within the studied repositories.  Since they are all Git repositories, one might expect the commit timestamps to be after the initial release of Git (around 2005).  It is however possible some repositories were in a different version control system (such as CVS or Subversion) and converted to Git.  For the sake of this study, we decided to investigate any commit timestamp prior to the release of CVS version 1.0 (19 November 1990).  The relevant part of the Boa query in \figref{fig:boaquery} is lines \ref{ln:old1}--\ref{ln:old2}.

The result of this query found 4,735 suspicious commit timestamps from 82 projects.  For those projects, this represents 3.45\% of their total commits.  For the full dataset, this represents 0.02\% of the commits.  In total, there were 23 unique suspicious timestamps (note: Boa timestamps are given as Unix timestamps with milliseconds), listed here along with the number of times they occurred and their conversion to a human readable date format:

\begin{lstlisting}[numbers=none]
$ cut -d'=' -f2 P.txt | sort -n | uniq
-2044178335000000#  1 time,  03/23/1905, 12:41:05 PM
0                #4677 times,01/01/1970, 12:00:00 AM
730000000        #  1 time,  01/01/1970, 12:12:10 AM
956000000        #  1 time,  01/01/1970, 12:15:56 AM
1585000000       #  1 time,  01/01/1970, 12:26:25 AM
1601000000       #  1 time,  01/01/1970, 12:26:41 AM
1627000000       #  1 time,  01/01/1970, 12:27:07 AM
3495000000       #  1 time,  01/01/1970, 12:58:15 AM
3523000000       #  1 time,  01/01/1970, 12:58:43 AM
7403000000       #  1 time,  01/01/1970, 02:03:23 AM
7558000000       #  1 time,  01/01/1970, 02:05:58 AM
7923000000       #  1 time,  01/01/1970, 02:12:03 AM
88210000000      #  1 time,  01/02/1970, 12:30:10 AM
88211000000      #  2 times, 01/02/1970, 12:30:11 AM
88212000000      #  3 times, 01/02/1970, 12:30:12 AM
88213000000      #  2 times, 01/02/1970, 12:30:13 AM
127771000000     #  1 time,  01/02/1970, 11:29:31 AM
179895000000     #  1 time,  01/03/1970, 01:58:15 AM
255447000000     #  1 time,  01/03/1970, 10:57:27 PM
1000000000000    # 25 times, 01/12/1970, 13:46:40 PM
315772873000000  #  1 time,  01/03/1980, 06:41:13 PM
566635987000000  #  5 times, 12/16/1987, 06:53:07 AM
589770257000000  #  5 times, 09/09/1988, 02:04:17 AM
\end{lstlisting}

As can be seen, the majority of the suspicious timestamps are the value 0.  There are however a handful of other suspicious timestamps.  For example, the 8 timestamps on January 2, 1970 at 12:30 all come from a single project that was ported over from Microsoft's CodePlex.\footnote{\url{https://github.com/KevinHoward/Irony}}  Most likely there was a problem in that porting process.

\begin{figure*}
\centering
\includegraphics[width=\linewidth]{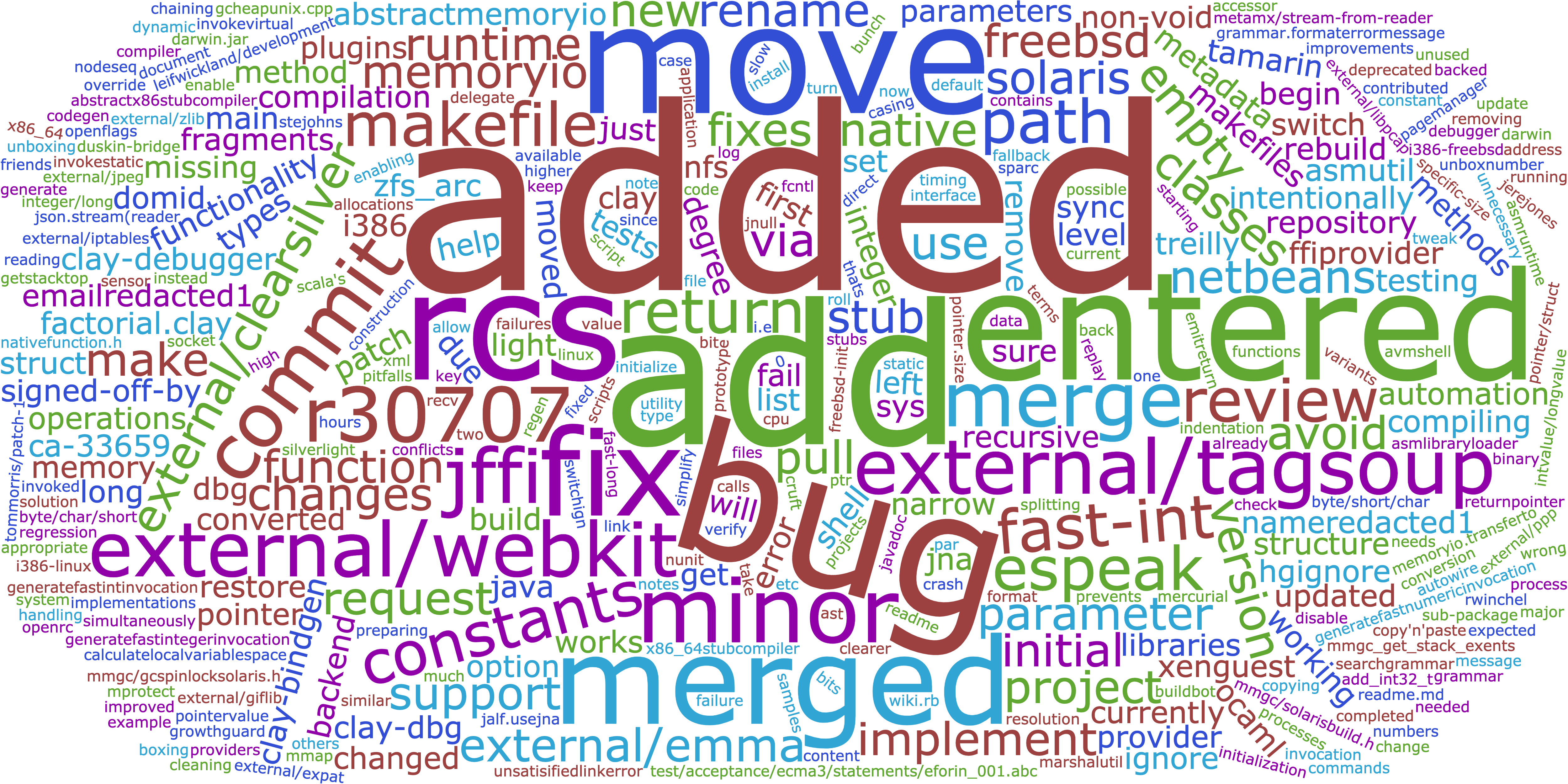}
\caption{Word Cloud of tokens appearing in suspicious commit messages, excluding commits containing the frequent term ``git-svn-id''.  English stop words were removed.}
\label{fig:wordcloud-old}
\end{figure*}

In fact, many of these suspicious timestamped commits seem to come from tools, such as \texttt{git-svn}.\footnote{\url{https://github.com/maodouzi/PY/commits/master}}  This tool was popular in the period between when Subversion was more common and people were starting to move to Git.  It allows maintaining a Git clone of a Subversion repository but required inserting `git-svn-id' tags into the commit messages to properly track the SVN repository.  We were able to verify 3,122 of the commit logs via GitHub's API, and 2,847 of those commits (91\%) contain a \texttt{git-svn-id} tag in the message.

Since that tool accounted for such a large portion of the commits, we investigated all the remaining commits by generating a word cloud of the commit logs, shown in \figref{fig:wordcloud-old}.  We note that several frequent words include 'external/' and probably indicate a repository tracking another (external) repository, most likely via some tool.

We also investigated dates that might be in the future. For this we used a cutoff time of the Boa dataset's release date (31 October 2019). Any commit with a time later than that release date was output.  This analysis yielded 11 commits from 3 projects where the dates were in the years 2025, 2027, and 2037.  Clearly these commits have invalid dates. A manual inspection of these commits showed the commits were (based on the Git graph) in between commits with dates that appear valid, indicating the years were off. Most likely these invalid dates were generated through either user error or misconfigured clocks.

\subsection{Finding Out-of-order Commits}
\label{subsec:order-commits}

Another possible problem with Git allowing users and tools to set the commit date is that the date specified might \textit{seem} valid, but actually be wrong.  This could lead to a graph where a particular node has a commit date that is actually older than its parent node.  Obviously such a case should not make sense.  This might be due to a misconfigured clock on a particular computer\footnote{\url{https://stackoverflow.com/questions/633353}}, specifying the wrong time zone\footnote{\url{https://stackoverflow.com/questions/52507279}}, or any other number of causes.\footnote{\url{https://stackoverflow.com/questions/16259105}}  We call these \textit{out-of-order commits}.

In this section, we investigate how frequently such out-of-order commits occur in Boa's dataset.  To do this, we traverse the revision list of each code repository in order and compare the commit date of a revision to the commit date of the previous revision.  Due to how Boa linearizes the commit graph using a topological sort, this might not technically be a parent (indeed, commit nodes might also have multiple parents due to branching) but it can give us insight into this problem.

In the first attempt at writing this query, we noticed a lot of results where one of the two commits were explicitly marked (in the log) as a merge commit.  We decided to filter those out as merging behavior might induce a lot of false positives.  The relevant part of the Boa query in \figref{fig:boaquery} is lines \ref{ln:order1}--\ref{ln:order2}.

Running this query gave us 26,252 suspicious commits from 4,744 projects.  For those projects, this represents 0.55\% of their total commits.  For the full dataset, this represents 0.11\% of the commits.  We used the GitHub API to download the JSON metadata for as many of the commits as possible and for any missing commits attempted to obtain JSON metadata from Software Heritage.  This left us with 25,956 commits, which we then verified were either older or newer than their parents.

Our verification process indicated that a total of 18,762 commits had at least one parent that was newer than the commit itself:

\begin{lstlisting}
$ grep ': BAD' order-verified.txt | wc -l
   18762
\end{lstlisting}

Inspecting a small sample of 20 of those commits, we see the time difference mostly runs under an hour (9 commits) or under one day (8 commits).  One commit was almost 19 days out of order.  Although we can't tell from the commits themselves, we suspect that bad server clocks and timezone issues account for most of these observed differences that are one day or less.  In the next sections we look at some common tools, users, and projects observed in the out-of-order dataset.

\begin{figure*}
\centering
\includegraphics[width=\linewidth]{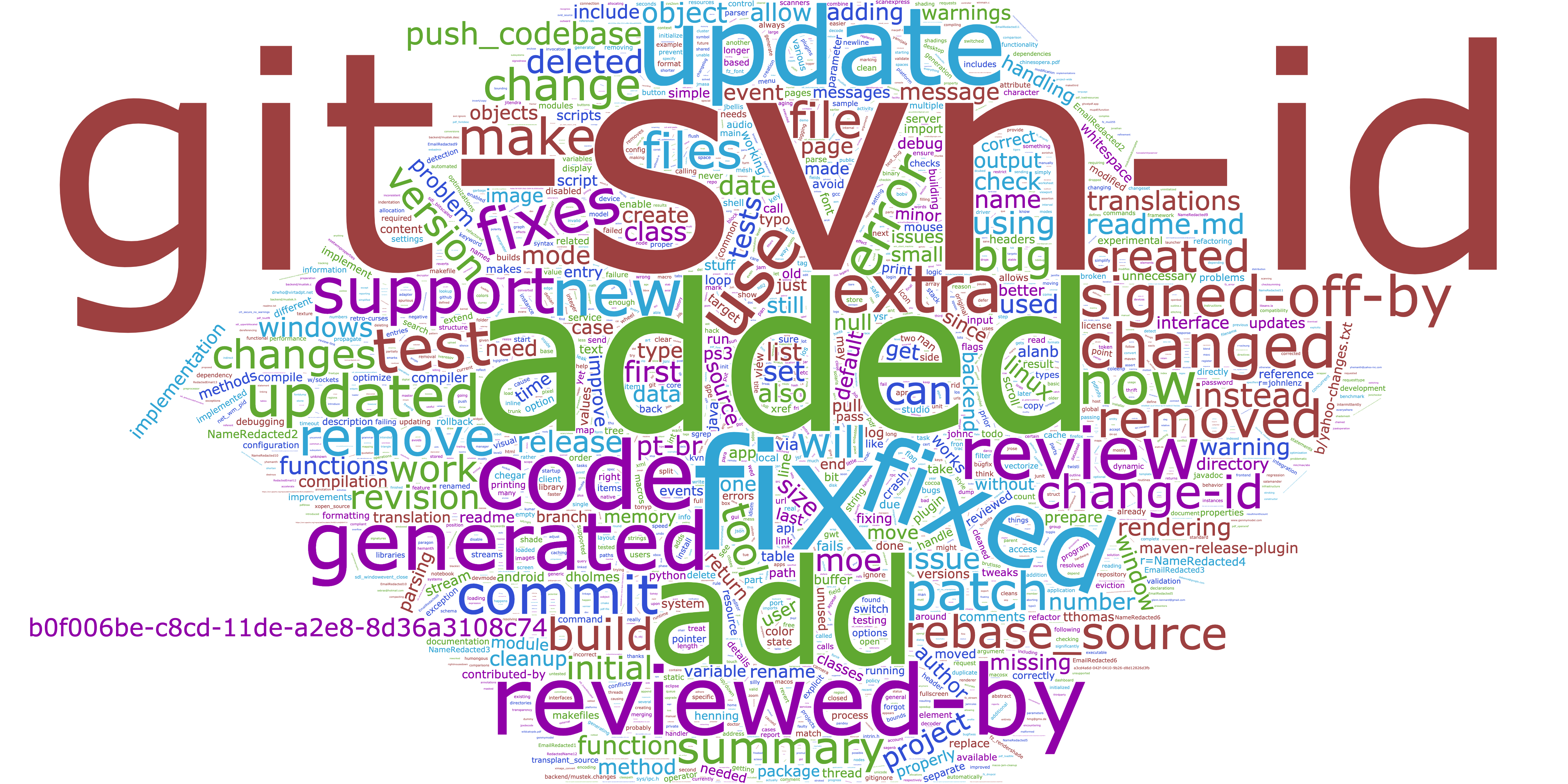}
\caption{Word Cloud of tokens appearing in bad commit messages.  English stop words were removed.}
\label{fig:wordcloud-order}
\end{figure*}

\subsubsection{Common Tools}

We further processed the commits suspected to be out of order.  Having done so, we collected all commit messages and removed English stop words to produce a word cloud (\figref{fig:wordcloud-order}) allowing us to visualise terms frequently used in the bad commits.  By doing so, we were able to note a handful of tools that have a tendency to produce bad commit timestamps.

\begin{lstlisting}
$ grep 'Reviewed-by' logs-order.txt | wc -l
984
$ grep 'Change-Id' logs-order.txt | wc -l
363

$ grep 'rebase_source' logs-order.txt | wc -l
390
$ grep -i '\bhg\b' logs-order.txt | wc -l
539

$ grep 'MOE' logs-order.txt | wc -l
465
$ grep 'push_codebase' logs-order.txt | wc -l
232
$ grep 'MOE\|push_codebase' logs-order.txt | wc -l
465
\end{lstlisting}

Review systems like Gerrit\footnote{\url{https://gerritcodereview.com}} seem to be a frequent contributor to bad commits, as found by the \texttt{Change-Id} commit footer (363 times).  We suspect this is due to the ``push, review, commit/rebase, force push, GOTO review'' method that is used by many participants in the code review process. 

We also frequently found other commit log footers, like \texttt{Reviewed-by} (984 times total).  These are used in other review processes, which involve either rebasing to edit commit messages to include them, or passing patch sets via email.

We also noticed another mixed VCS, namely \texttt{hg-git}\footnote{\url{https://www.mercurial-scm.org/wiki/HgGit}}, which allows a Mercurial user to manipulate a Git repository using Mercurial commands.  In particular, we note the addition of metadata to commits, with the \texttt{rebase\_source} footer (seen 390 times), which is likely a result of rebases on Git repositories using Mercurial or similar tools.  Mercurial's abbreviation, \texttt{hg} is also found 539 times.

Google produced a tool, MOE\footnote{\url{https://opensource.google/projects/moe}} (Make Open Easy) which is used to synchronize two repositories, one internal, and one open to the public.  This tool can synchronize, translate content between kinds of repositories, and scrub content from a repository.  Because of these features, we suspect that use of this tool produced a mismatch between repositories, where an open-source repository received patches from an internal repository after receiving patches from other contributors.  We see MOE related messages show up 465 times across the bad commits.


\subsubsection{Common Users}

Using the collected commit data, we analyzed commit author information and counted the number of commits made by the top 20 contributors of the bad commits in our dataset.  From this, we found that 5,279 commits (28\%) were made by the top 20 users (with all `(no name)' users considered as one).

\subsubsection{Common Projects}

Similarly, we collected the project each bad commit belonged to and found that the top 20 projects contributed 3,972 of the bad commits (21\%), see \tabref{tbl:proj-faulty}.

\begin{table}
\caption{Top 20 projects with the most faulty commits.}
\label{tbl:proj-faulty}
\centering
\begin{tabular}{rl}
\hline
\textbf{Bad Commits} & \textbf{Project} \\
\hline
584 & glob3mobile/g3m \\
420 & mdcurtis/micromanager-upstream \\
407 & openspim/micromanager \\
333 & axDev-JDK/jdk \\
317 & GrammaticalFramework/GF \\
198 & mytskine/mupdf-unofficial \\
187 & oferfrid/PrecisExciteTCP \\
166 & axDev-JDK/hotspot \\
161 & uditrugman/openjdk8-jdk \\
132 & uditrugman/openjdk8-hotspot \\
113 & xapi-project/xen-api \\
101 & zeniko/mupdf \\
101 & muennich/mupdf \\
101 & michaelcadilhac/pdfannot \\
101 & lmurmann/mupdf \\
101 & libretro/RetroArch \\
101 & issuestand/mupdf \\
101 & crow-misia/mupdf \\
101 & clchiou/mupdf \\
101 & ccxvii/mupdf \\
\hline
\end{tabular}

\end{table}

Note that several of the projects appear to be clones, where the original repository most likely contained some bad commit timestamps.  Boa only contains non-forked repositories, so these projects most likely cloned and uploaded the repository without utilizing GitHub's fork feature.

\subsection{Summary}

To summarize, we were able to find thousands of bad commit timestamps.  Many of these commits seem to originate from tool use, especially tools that migrate or synchronize between two version control systems.  In addition, we saw a small number of users and projects seem to contribute a large number of the bad commit timestamps.
\section{Threats to Validity}

The keyword-based search technique to identify MSR papers using time-based data is sound, as we manually verified the results.  It is however not complete, as a paper might have utilized time-based data without using any of the selected keywords.  Thus the 35\% of papers identified is a lower bound and the actual number of papers utilizing time-based data might be even higher.

The use of Boa and the existence of cloned repositories in its dataset is a potential further threat.  In those commits which were considered suspiciously old, there were 3,412 unique commit ids, with 821 of these ids repeated (22.7\%).  In those which were out of order, there was a total of 18,685 unique commit ids, with 9,726 repeated (52\%).  However, due to the method by which commits were examined (selecting by ID alone from either GitHub or the Software Heritage archive) further analysis on commit contents was based on on the unique commits.  This underscores the necessity of careful de-duplication (relying on things other than just GitHub's fork feature to record the creation of forks) as well as serves to underscore the wide-range of faulty commits.

Note that 869 out of the 4,785 projects (18.2\%) identified with time data problems no longer exist on GitHub (as of 28 December 2020).  These projects are however still in the released Boa dataset, so we maintained their results in this study.  We attempted to mitigate this threat by validating the data directly with GitHub using their API, and for the projects that were missing we utilized the Software Heritage Archive.  Note however that the times Boa and Software Heritage indexed the projects might differ, and thus there were some commits found by Boa that we were not able to verify. Such commits were removed from the dataset.  Of the 29,896 total commits found by Boa, 25,012 (83.66\%) were still on GitHub, 3,568 (11.93\%) were found on Software Heritage, and 1,316 (4.40\%) were excluded.  We do not believe this is a problem however, as the point of the analysis was to see if bad time data exists, not to fully account for all such cases.

Some of the commits might actually suffer from multiple problems.  To quantify how many commits have a bad timestamp and also are out of order, we intersect the two results:

\begin{lstlisting}
$ wc -l *-output.txt
    4735 old-output.txt
   26252 order-output.txt
   30987 total
$ comm old-output.txt order-output.txt | wc -l
   29514
$ comm -12 old-output.txt order-output.txt | wc -l
    1473
\end{lstlisting}

\noindent This shows there are 1,473 (4.99\%) commits in Boa's dataset that potentially suffered from both the out-of-order error as well as being suspiciously old.  Most of those (all but 11) are timestamps of 0.

A threat to external validity is that we only studied Git repositories. However, although we did not quantify it explicitly, Subversion also allows developers to modify (and even remove!) the commit date: \texttt{svn propset -rXXX --revprop svn:date ...}.  Running the same Boa queries on Boa's older dataset from SourceForge (which contains Subversion repositories) shows there are both suspiciously old\footnote{\url{http://boa.cs.iastate.edu/boa/?q=boa/job/public/91547}} and out-of-order\footnote{\url{http://boa.cs.iastate.edu/boa/?q=boa/job/public/91552}} commits in that data too.  The results however may not generalize to any VCS that disallows modifying commit timestamps.
\section{Discussion}
\label{sec:discussion}

We showed that time-based data is utilized by a large number of MSR research (at least 35\% of papers).  We then described some possible problems with timestamps in Git data, the most used data kind, and attempted to quantify how often those problems occur in the most used data source, GitHub.  In this section, we discuss some guidelines for handling time-based data.


When order is a component of an analysis, handling suspicious commits is recommended.  To do this, we recommend that any commit with a timestamp less than 1 is removed.  For the data we observed, this filter would remove about 98\% of suspicious commits.

In general, we also recommend searching the commit logs for projects that contain a `\texttt{git-svn-id}' tag.


To handle the problem of out-of-order commits, we recommend four strategies: \begin{enumerate*}
\item filtering commits before a certain date;
\item filtering commits belonging to certain projects;
\item filtering only commits which are out-of-order; and/or,
\item using a robust method of mining commits with rebasing.
\end{enumerate*} We will discuss each of these in turn, including the benefits and the potential problems each brings to the table.

\subsubsection{Filtering Before a Specific Date}

\begin{table}
\caption{Percent of faulty commits removed by filtering commits from or before a given year.}
\label{tbl:filtering-by-year}
\centering
\begin{tabular}{cr}
\hline
\textbf{Filter Date} & \textbf{Percent Removed} \\
\hline
$\leq 2015$ & 100.00\% \\
$\leq 2014$ & 99.85\% \\
$\leq 2013$ & 98.28\% \\
$\leq 2012$ & 70.93\% \\
$\leq 2011$ & 51.15\% \\
$\leq 2010$ & 36.72\% \\
$\leq 2009$ & 25.41\% \\
$\leq 2008$ & 18.69\% \\
$\leq 2007$ & 13.39\% \\
$\leq 2006$ & 11.60\% \\
$\leq 2005$ & 10.58\% \\
$\leq 2004$ & 9.50\% \\
$\leq 2003$ & 8.70\% \\
$\leq 2002$ & 8.66\% \\
$\leq 2001$ & 8.52\% \\
$\leq 2000$ & 8.37\% \\
$\leq 1999$ & 8.29\% \\
$\leq 1998$ & 8.28\% \\
$\leq 1997$ & 8.25\% \\
$\leq 1996$ & 8.20\% \\
$\leq 1994$ & 8.19\% \\
$\leq 1992$ & 7.18\% \\
$\leq 1970$ & 7.18\% \\
\hline
\end{tabular}

\end{table}

The first filtering method we suggest is removing commits from \emph{before} a specific date.  We suggest this method due to its relative simplicity, as well as its effect (see \tabref{tbl:filtering-by-year}).  In particular, we suggest removing all commits before 1 January 2014, as doing so could remove 98.28\% of all out-of-order commits.  Even filtering this much data still leaves (at the time of writing) seven years of historical data to study.

If the research question requires longer history, this method may not be feasible and one of the other two methods is recommended.

\subsubsection{Filtering Specific Projects}

Given the exceedingly wide range of projects available on GitHub and similar sites, the removal of projects known to have a large number of out-of-order commits still leaves a large available corpus.  The benefit is that a longer history can be maintained.  This is a bit more complicated than simply filtering by a specific date though, as a list of so called ``bad'' projects would need to be known.  The MSR community could work toward maintaining such a list.

\subsubsection{Filtering only Out-of-Order Commits}

This filtering method is the most complex, requiring each commit to be examined in turn, however, this method has the benefit of not removing other history, and could be used to enable study of a repository that otherwise may be problematic, or has substantial pre-2014 history.  Further, this method has been used previously (e.g., \cite{steff12}).



\subsubsection{Ordering on Author Date}

Git commits store two timestamps: the \textbf{author date} and the \textbf{committer date}.  The author date is the date the commit was originally made.  The committer date is essentially the last date the commit was modified.  Certain commands, e.g. cherry picking and rebasing, modify the commits and thus will change the committer date.

We recommend most studies utilize the author date instead of committer date.  Most of the time, the two dates are identical.  However, for repositories with rebase-heavy workflows, such as those utilizing Gerrit and similar review platforms, techniques such as those recommended by \textcite{paixao19:rebasing} might be required.


\subsubsection{Summary}

The exact method(s) utilized will depend highly on the specific research questions being answered.  For example, if the research questions require a long history then filtering by date might not be the best approach.  Additionally, researchers need to decide if it is acceptable to simply drop the bad commits alone, or if a project with bad commits should be excluded entirely.
\section{Conclusion}
\label{sec:conclusion}

The use of time-based data in MSR studies is wide-spread, in at least 35\% of MSR papers.  Properly handling this time-based data is thus very important.  However, the diversity of tools and workflows used to generate the time-based data can present challenges.  In particular, ensuring that time data is consistent and maintains linearity is important.  Further, we have found that many papers do not describe cleaning or filtering of time data, with those that do describe filtering tending towards simple techniques like selection from a defined time span or selection of data before a certain date.  Some papers have used more robust or rigorous techniques, and may thus avoid some of the time-related problems found in data such as Git repositories.

To remedy potential time-based issues in VCS data, such as that coming from GitHub, we recommend a simple filter to drop any timestamp less than 1 as well as a more complex filtering to remove out-of-order commits.  Ideally, each repository would be analyzed to detect and remove the out-of-order commits, but depending on the need a simple cutoff filter removing commits prior to 2014 might suffice.  Applying both filters (the second filter actually implies the first) is very simple and would remove around 98\% of all observed bad commits.

In the future, we would like to sample some datasets from prior publications to quantify how often bad time-based data occurred in those studies and quantify how it may have affected the published results.  We would also like to investigate potential problems in other kinds of time data, such as issue reports.  We would also like to investigate how time-based data is used when training machine learning models and if issues arise from training on later observed data and then classifying on older data.

\section*{Acknowledgment}
The authors would like to thank Yijia Huang, Tien N. Nguyen, and Hridesh Rajan for insightful discussions that inspired this paper.  We also thank the anonymous reviewers for several suggestions that have substantially improved this paper.

\printbibliography

@dataset{replication,
  author       = {Flint, Samuel W. and
                  Chauhan, Jigyasa and
                  Dyer, Robert},
  title        = {{Replication package for "Escaping the Time Pit: 
                   Pitfalls and Guidelines for Using Time-Based Git
                   Data"}},
  month        = mar,
  year         = 2021,
  publisher    = {Zenodo},
  version      = {1.1.0},
  doi          = {10.5281/zenodo.4625288},
  url          = {https://doi.org/10.5281/zenodo.4625288}
}

@inproceedings{boa,
 author = {Dyer, Robert and Nguyen, Hoan Anh and Rajan, Hridesh and Nguyen, Tien N.},
 title = {Boa: A Language and Infrastructure for Analyzing Ultra-large-scale Software Repositories},
 booktitle = {Proceedings of the 2013 International Conference on Software Engineering},
 series = {ICSE '13},
 year = {2013},
 pages = {422--431},
 doi = {10.5555/2486788.2486844},
 publisher = {IEEE Press},
}

@misc{boa-website,
  title = {Boa: Mining Ultra-Large-Scale Software Repositories},
  author = {Dyer, Robert and Nguyen, Hoan Anh and Rajan, Hridesh and Nguyen, Tien N.},
  howpublished = {\url{http://boa.cs.iastate.edu/}},
  note = {Accessed: 2020-12-28},
  year = {2020},
}

@misc{software-heritage-archive,
  title = {Software Heritage Archive},
  author = {The Software Heritage developers},
  howpublished = {\url{https://archive.softwareheritage.org/}},
  note = {Accessed: 2020-12-28},
  year = {2020},
}

@INPROCEEDINGS{robles10,
  author = {G. {Robles}},
  booktitle = {7th IEEE Working Conference on Mining Software Repositories},
  title={Replicating {MSR}: A study of the potential replicability of papers published in the {Mining Software Repositories} proceedings},
  series = {MSR '10},
  year = {2010},
  pages = {171-180},
  doi = {10.1109/MSR.2010.5463348},
}

@inproceedings{kotti19,
author = {Kotti, Zoe and Spinellis, Diomidis},
title = {Standing on Shoulders or Feet? The Usage of the {MSR} Data Papers},
year = {2019},
publisher = {IEEE Press},
%url = {https://doi.org/10.1109/MSR.2019.00085},
doi = {10.1109/MSR.2019.00085},
booktitle = {Proceedings of the 16th International Conference on Mining Software Repositories},
pages = {565–576},
numpages = {12},
keywords = {bibliometrics, reproducibility, data paper, software engineering data, data showcase track},
location = {Montreal, Quebec, Canada},
series = {MSR '19}
}

@INPROCEEDINGS{cosentino16,
  author={V. {Cosentino} and J. L. C. {Izquierdo} and J. {Cabot}},
  booktitle={2016 IEEE/ACM 13th Working Conference on Mining Software Repositories (MSR)}, 
  title={Findings from {GitHub}: Methods, Datasets and Limitations}, 
  year={2016},
  volume={},
  number={},
  pages={137-141},
  doi={}}

@inproceedings{hemmati13,
author = {Hemmati, Hadi and Nadi, Sarah and Baysal, Olga and Kononenko, Oleksii and Wang, Wei and Holmes, Reid and Godfrey, Michael W.},
title = {The {MSR} Cookbook: Mining a Decade of Research},
year = {2013},
isbn = {9781467329361},
publisher = {IEEE Press},
pages = {343–352},
numpages = {10},
location = {San Francisco, CA, USA},
series = {MSR '13}
}

@inproceedings{demeyer13,
author = {Demeyer, Serge and Murgia, Alessandro and Wyckmans, Kevin and Lamkanfi, Ahmed},
title = {Happy Birthday! A Trend Analysis on Past {MSR} Papers},
year = {2013},
isbn = {9781467329361},
publisher = {IEEE Press},
booktitle = {Proceedings of the 10th Working Conference on Mining Software Repositories},
pages = {353–362},
numpages = {10},
location = {San Francisco, CA, USA},
series = {MSR '13}
}

@inproceedings{ghezzi13,
author = {Ghezzi, Giacomo and Gall, Harald C.},
title = {Replicating Mining Studies with {SOFAS}},
year = {2013},
isbn = {9781467329361},
publisher = {IEEE Press},
booktitle = {Proceedings of the 10th Working Conference on Mining Software Repositories},
pages = {363–372},
numpages = {10},
location = {San Francisco, CA, USA},
series = {MSR '13}
}

@INPROCEEDINGS{bird09,
  author={C. {Bird} and P. C. {Rigby} and E. T. {Barr} and D. J. {Hamilton} and D. M. {German} and P. {Devanbu}},
  booktitle={2009 6th IEEE International Working Conference on Mining Software Repositories}, 
  title={The promises and perils of mining git}, 
  year={2009},
  volume={},
  number={},
  pages={1-10},
  doi={10.1109/MSR.2009.5069475}}

@inproceedings{liu2020,
author = {Liu, Yalin and Lin, Jinfeng and Cleland-Huang, Jane},
title = {Traceability Support for Multi-Lingual Software Projects},
year = {2020},
%isbn = {9781450375177},
publisher = {Association for Computing Machinery},
address = {New York, NY, USA},
%url = {https://doi.org/10.1145/3379597.3387440},
doi = {10.1145/3379597.3387440},
booktitle = {Proceedings of the 17th International Conference on Mining Software Repositories},
pages = {443–454},
numpages = {12},
keywords = {Generalized Vector Space Model, Traceability, Cross-lingual information retrieval},
location = {Seoul, Republic of Korea},
series = {MSR '20}
}

@inproceedings{kikas16,
author = {Kikas, Riivo and Dumas, Marlon and Pfahl, Dietmar},
title = {Using Dynamic and Contextual Features to Predict Issue Lifetime in {GitHub} Projects},
year = {2016},
%isbn = {9781450341868},
publisher = {Association for Computing Machinery},
address = {New York, NY, USA},
%url = {https://doi.org/10.1145/2901739.2901751},
doi = {10.1145/2901739.2901751},
booktitle = {Proceedings of the 13th International Conference on Mining Software Repositories},
pages = {291–302},
numpages = {12},
keywords = {issue lifetime prediction, issue tracking, mining software repositories},
location = {Austin, Texas},
series = {MSR '16}
}

@inproceedings{software-heritage,
  title = {{Software Heritage}: Why and How to Preserve Software Source Code},
  author = {Roberto Di Cosmo and Stefano Zacchiroli},
%  url = {https://hal.archives-ouvertes.fr/hal-01590958},
  year = {2017},
  date = {2017-09-25},
  booktitle = {iPRES 2017: 14th International Conference on Digital Preservation},
  address = {Kyoto, Japan},
}

@inproceedings{steff12,
author = {Steff, Maximilian and Russo, Barbara},
title = {Co-Evolution of Logical Couplings and Commits for Defect Estimation},
year = {2012},
isbn = {9781467317610},
publisher = {IEEE Press},
booktitle = {Proceedings of the 9th IEEE Working Conference on Mining Software Repositories},
pages = {213–216},
numpages = {4},
location = {Zurich, Switzerland},
series = {MSR '12}
}

@inproceedings{claes20,
author = {Claes, Ma\"{e}lick and M\"{a}ntyl\"{a}, Mika V.},
title = {{20-MAD}: 20 Years of Issues and Commits of {Mozilla} and {Apache} Development},
year = {2020},
%isbn = {9781450375177},
publisher = {Association for Computing Machinery},
address = {New York, NY, USA},
%url = {https://doi.org/10.1145/3379597.3387487},
doi = {10.1145/3379597.3387487},
booktitle = {Proceedings of the 17th International Conference on Mining Software Repositories},
pages = {503–507},
numpages = {5},
location = {Seoul, Republic of Korea},
series = {MSR '20}
}

@inproceedings{xu18,
author = {Xu, Yulin and Zhou, Minghui},
title = {A Multi-Level Dataset of {Linux} Kernel Patchwork},
year = {2018},
%isbn = {9781450357166},
publisher = {Association for Computing Machinery},
address = {New York, NY, USA},
%url = {https://doi.org/10.1145/3196398.3196475},
doi = {10.1145/3196398.3196475},
booktitle = {Proceedings of the 15th International Conference on Mining Software Repositories},
pages = {54–57},
numpages = {4},
location = {Gothenburg, Sweden},
series = {MSR '18}
}

@INPROCEEDINGS{baysal12,
  author={O. {Baysal} and R. {Holmes} and M. W. {Godfrey}},
  booktitle={2012 9th IEEE Working Conference on Mining Software Repositories (MSR)}, 
  title={Mining usage data and development artifacts}, 
  year={2012},
  volume={},
  number={},
  pages={98-107},
  doi={10.1109/MSR.2012.6224305}}

@inproceedings{gonzalez-barahona15,
author = {Gonzalez-Barahona, Jesus M. and Robles, Gregorio and Izquierdo-Cortazar, Daniel},
title = {The {MetricsGrimoire} Database Collection},
year = {2015},
isbn = {9780769555942},
publisher = {IEEE Press},
booktitle = {Proceedings of the 12th Working Conference on Mining Software Repositories},
pages = {478–481},
numpages = {4},
location = {Florence, Italy},
series = {MSR '15}
}

@inproceedings{robles14,
author = {Robles, Gregorio and Gonz\'{a}lez-Barahona, Jes\'{u}s M. and Cervig\'{o}n, Carlos and Capiluppi, Andrea and Izquierdo-Cort\'{a}zar, Daniel},
title = {Estimating Development Effort in Free/Open Source Software Projects by Mining Software Repositories: A Case Study of {OpenStack}},
year = {2014},
%isbn = {9781450328630},
publisher = {Association for Computing Machinery},
address = {New York, NY, USA},
%url = {https://doi.org/10.1145/2597073.2597107},
doi = {10.1145/2597073.2597107},
booktitle = {Proceedings of the 11th Working Conference on Mining Software Repositories},
pages = {222–231},
numpages = {10},
keywords = {open source, free software, mining software repositories, Effort estimation},
location = {Hyderabad, India},
series = {MSR 2014}
}

@INPROCEEDINGS{goeminne13,
  author={M. {Goeminne} and M. {Claes} and T. {Mens}},
  booktitle={2013 10th Working Conference on Mining Software Repositories (MSR)},
  title={A historical dataset for the {Gnome} ecosystem},
  year={2013},
  volume={},
  number={},
  pages={225-228},
  doi={10.1109/MSR.2013.6624032}}

@inproceedings{sadowski11,
author = {Sadowski, Caitlin and Lewis, Chris and Lin, Zhongpeng and Zhu, Xiaoyan and Whitehead, E. James},
title = {An Empirical Analysis of the {FixCache} Algorithm},
year = {2011},
%isbn = {9781450305747},
publisher = {Association for Computing Machinery},
address = {New York, NY, USA},
%url = {https://doi.org/10.1145/1985441.1985475},
doi = {10.1145/1985441.1985475},
booktitle = {Proceedings of the 8th Working Conference on Mining Software Repositories},
pages = {219–222},
numpages = {4},
keywords = {software bug prediction, fixcache},
location = {Waikiki, Honolulu, HI, USA},
series = {MSR '11}
}

@inproceedings{durieux20,
author = {Durieux, Thomas and Le Goues, Claire and Hilton, Michael and Abreu, Rui},
title = {Empirical Study of Restarted and Flaky Builds on {Travis} {CI}},
year = {2020},
%isbn = {9781450375177},
publisher = {Association for Computing Machinery},
address = {New York, NY, USA},
%url = {https://doi.org/10.1145/3379597.3387460},
doi = {10.1145/3379597.3387460},
booktitle = {Proceedings of the 17th International Conference on Mining Software Repositories},
pages = {254–264},
numpages = {11},
location = {Seoul, Republic of Korea},
series = {MSR '20}
}

@inproceedings{hayashi19,
author = {Hayashi, Junichi and Higo, Yoshiki and Matsumoto, Shinsuke and Kusumoto, Shinji},
title = {Impacts of Daylight Saving Time on Software Development},
year = {2019},
publisher = {IEEE Press},
%url = {https://doi.org/10.1109/MSR.2019.00076},
doi = {10.1109/MSR.2019.00076},
booktitle = {Proceedings of the 16th International Conference on Mining Software Repositories},
pages = {502–506},
numpages = {5},
location = {Montreal, Quebec, Canada},
series = {MSR '19}
}

@inproceedings{pimentel19,
author = {Pimentel, Jo\~{a}o Felipe and Murta, Leonardo and Braganholo, Vanessa and Freire, Juliana},
title = {A Large-Scale Study about Quality and Reproducibility of {Jupyter} Notebooks},
year = {2019},
publisher = {IEEE Press},
%url = {https://doi.org/10.1109/MSR.2019.00077},
doi = {10.1109/MSR.2019.00077},
booktitle = {Proceedings of the 16th International Conference on Mining Software Repositories},
pages = {507–517},
numpages = {11},
location = {Montreal, Quebec, Canada},
series = {MSR '19}
}

@inproceedings{ahasanuzzaman16,
author = {Ahasanuzzaman, Muhammad and Asaduzzaman, Muhammad and Roy, Chanchal K. and Schneider, Kevin A.},
title = {Mining Duplicate Questions in {Stack} {Overflow}},
year = {2016},
%isbn = {9781450341868},
publisher = {Association for Computing Machinery},
address = {New York, NY, USA},
%url = {https://doi.org/10.1145/2901739.2901770},
doi = {10.1145/2901739.2901770},
booktitle = {Proceedings of the 13th International Conference on Mining Software Repositories},
pages = {402–412},
numpages = {11},
location = {Austin, Texas},
series = {MSR '16}
}

@inproceedings{antoniol05,
author = {Antoniol, Giuliano and Rollo, Vincenzo Fabio and Venturi, Gabriele},
title = {Linear Predictive Coding and Cepstrum Coefficients for Mining Time Variant Information from Software Repositories},
year = {2005},
%isbn = {1595931236},
publisher = {Association for Computing Machinery},
address = {New York, NY, USA},
%url = {https://doi.org/10.1145/1083142.1083156},
doi = {10.1145/1083142.1083156},
booktitle = {Proceedings of the 2005 International Workshop on Mining Software Repositories},
pages = {1–5},
numpages = {5},
keywords = {data mining, software evolution},
location = {St. Louis, Missouri},
series = {MSR '05}
}

@inproceedings{walker06,
author = {Walker, Robert J. and Holmes, Reid and Hedgeland, Ian and Kapur, Puneet and Smith, Andrew},
title = {A Lightweight Approach to Technical Risk Estimation via Probabilistic Impact Analysis},
year = {2006},
%isbn = {1595933972},
publisher = {Association for Computing Machinery},
address = {New York, NY, USA},
%url = {https://doi.org/10.1145/1137983.1138008},
doi = {10.1145/1137983.1138008},
booktitle = {Proceedings of the 2006 International Workshop on Mining Software Repositories},
pages = {98–104},
numpages = {7},
keywords = {decision support, probabilistic impact analysis, revision history, technical risk estimation},
location = {Shanghai, China},
series = {MSR '06}
}

@inproceedings{kagdi06,
author = {Kagdi, Huzefa and Yusuf, Shehnaaz and Maletic, Jonathan I.},
title = {Mining Sequences of Changed-Files from Version Histories},
year = {2006},
%isbn = {1595933972},
publisher = {Association for Computing Machinery},
address = {New York, NY, USA},
%url = {https://doi.org/10.1145/1137983.1137996},
doi = {10.1145/1137983.1137996},
booktitle = {Proceedings of the 2006 International Workshop on Mining Software Repositories},
pages = {47–53},
numpages = {7},
keywords = {heuristics, change sequences, mining software repositories},
location = {Shanghai, China},
series = {MSR '06}
}

@INPROCEEDINGS{dambros10,
  author={M. {D'Ambros} and M. {Lanza} and R. {Robbes}},
  booktitle={2010 7th IEEE Working Conference on Mining Software Repositories (MSR 2010)}, 
  title={An extensive comparison of bug prediction approaches}, 
  year={2010},
  volume={},
  number={},
  pages={31-41},
  doi={10.1109/MSR.2010.5463279}}

@inproceedings{zimmerman04,
author = {Zimmermann, Thomas and Wei{\ss}gerber, Peter},
title = {Preprocessing {CVS} Data for Fine-Grained Analysis},
year = {2004},
pages = {2--6},
booktitle = {Proceedings of the 1st International Workshop on Mining Software Repositories},
location = {Edinburgh, Scotland},
series = {MSR '04}
}

@inproceedings{Wang_2020,
	doi = {10.1145/3379597.3387464},
	%url = {https://doi.org/10.1145%2F3379597.3387464},
	year = 2020,
	publisher = {{ACM}},
	author = {Peipei Wang and Chris Brown and Jamie A. Jennings and Kathryn T. Stolee},
	title = {An Empirical Study on Regular Expression Bugs},
	booktitle = {Proceedings of the 17th International Conference on Mining Software Repositories}
}

@inproceedings{Karampatsis_2020,
	doi = {10.1145/3379597.3387491},
	%url = {https://doi.org/10.1145%2F3379597.3387491},
	year = 2020,
	publisher = {{ACM}},
	author = {Rafael-Michael Karampatsis and Charles Sutton},
	title = {How Often Do Single-Statement Bugs Occur?},
	booktitle = {Proceedings of the 17th International Conference on Mining Software Repositories}
}

@inproceedings{Zhu_2019,
	doi = {10.1109/msr.2019.00068},
	%url = {https://doi.org/10.1109%2Fmsr.2019.00068},
	year = 2019,
	publisher = {{IEEE}},
	author = {Jiaxin Zhu and Jun Wei},
	title = {An Empirical Study of Multiple Names and Email Addresses in {OSS} Version Control Repositories},
	booktitle = {2019 {IEEE}/{ACM} 16th International Conference on Mining Software Repositories ({MSR})}
}

@inproceedings{Cito_2017,
	doi = {10.1109/msr.2017.67},
	%url = {https://doi.org/10.1109%2Fmsr.2017.67},
	year = 2017,
	publisher = {{IEEE}},
	author = {Jurgen Cito and Gerald Schermann and John Erik Wittern and Philipp Leitner and Sali Zumberi and Harald C. Gall},
	title = {An Empirical Analysis of the {Docker} Container Ecosystem on {GitHub}},
	booktitle = {2017 {IEEE}/{ACM} 14th International Conference on Mining Software Repositories ({MSR})}
}

@Inproceedings{gasser04,
  year = 2004,
  author = {Gasser, Les and Ripoche, Gabriel and Sandusky, Robert J.},
  title = {Research Infrastructure for Empirical Science of {F/OSS}},
  booktitle = {Proceedings of the 1st International Workshop on Mining Software Repositories},
}

@inproceedings{perils,
author = {Kalliamvakou, Eirini and Gousios, Georgios and Blincoe, Kelly and Singer, Leif and German, Daniel M. and Damian, Daniela},
title = {The Promises and Perils of Mining GitHub},
year = {2014},
isbn = {9781450328630},
publisher = {Association for Computing Machinery},
address = {New York, NY, USA},
url = {https://doi.org/10.1145/2597073.2597074},
doi = {10.1145/2597073.2597074},
booktitle = {Proceedings of the 11th Working Conference on Mining Software Repositories},
pages = {92--101},
numpages = {10},
location = {Hyderabad, India},
series = {MSR 2014}
}

@article{perils2,
author = {Kalliamvakou, Eirini and Gousios, Georgios and Blincoe, Kelly and Singer, Leif and German, Daniel M. and Damian, Daniela},
title = {An In-Depth Study of the Promises and Perils of Mining GitHub},
year = {2016},
issue_date = {October 2016},
publisher = {Kluwer Academic Publishers},
address = {USA},
volume = {21},
number = {5},
issn = {1382-3256},
url = {https://doi.org/10.1007/s10664-015-9393-5},
doi = {10.1007/s10664-015-9393-5},
journal = {Empirical Softw. Engg.},
month = oct,
pages = {2035–2071},
numpages = {37}
}

@InProceedings{paixao19:rebasing,
title={Rebasing in Code Review Considered Harmful: A Large-scale Empirical Investigation},
author = {Paixao, Matheus and Maia, Paulo Henrique},
years = {2019},
booktitle = {2019 19th International Working Conference on Source Code Analysis and Manipulation (SCAM)},
location = {Cleveland, OH, USA},
pages = {45--55},
doi = {10.1109/SCAM.2019.00014}
}

\end{document}